\documentclass[epj]{webofc}
\usepackage[varg]{txfonts}
\usepackage{lineno}
\usepackage{hyperref}
\usepackage{xcolor}
\usepackage{amsmath,amssymb}
\usepackage{booktabs}
\usepackage{caption}

\begin{document}

\title{The quest for dark matter in dwarf spheroidal galaxies with the Cherenkov Telescope Array}

\author{Francesco G. Saturni\inst{1,2}\fnsep\thanks{\email{francesco.saturni@inaf.it}} \and
        Gonzalo Rodr{\'i}guez-Fern{\'a}ndez\inst{3} \and
        Aldo Morselli\inst{3}~for the CTA Consortium
}

\institute{
          INAF -- Osservatorio Astronomico di Roma, Via Frascati 33, I-00040 Monte Porzio Catone (RM), Italy
\and
           ASI -- Space Science Data Center, Via del Politecnico snc, I-00133 Roma, Italy
\and
           INFN -- Sezione di Roma Tor Vergata, Via della Ricerca Scientifica 1, I-00133 Roma, Italy
          }

\abstract{
  Dwarf spheroidal galaxies are among the best environments that can be studied with Cherenkov telescopes for indirect searches of $\gamma$-ray signals coming from dark matter self-interaction (annihilation or decay), due to their proximity and negligible background emission. We present new determinations of the dark-matter amount -- i.e. the astrophysical factors $J$ and $D$ -- in dwarf-galaxy halos obtained through the MCMC Jeans analysis of their brightness and kinematic data. Such factors are of great importance to test the performances of the next-generation $\gamma$-ray instruments such as the Cherenkov Telescope Array in detecting dark-matter signals from astronomical environments, or constraining the limits to dark-matter physics parameters (particle mass and lifetime, annihilation cross section).
}

\maketitle

\section{Introduction}
\label{intro}
The problem of establishing the nature of Dark Matter (DM; \cite{Zwi33}) is one of the major open challenges in modern astrophysics. Several efforts, both on the side of elementary particles (e.g. Weakly Interacting Massive Particles, WIMPs; \cite{Apr18}) and macroscopic objects (Massive Compact Halo Objects, MACHOs; \cite{Tis07}), have been made to identify plausible DM candidates. However, the parameter space covered by such candidates ranges over several orders of magnitude in masses and cross sections \cite{Ros04}. The current framework for the astronomical searches for DM signals is based on the possibility that DM particles self-interact via annihilation or decay to produce final-state $\gamma$-ray photons \cite{Cir11} whose detection is potentially at reach of next-generation Cherenkov telescopes such as the Cherenkov Telescope Array (CTA; \cite{Cta17}). Here we present new estimates of the DM content in the dwarf spheroidal galaxies (dSphs; \cite{Str08}), which appear to be the best observational targets for indirect DM searches in terms of proximity and lack of background emission \cite{Eva04,Dor13}. In particular, we focus our analysis on the classical dSph Draco I (Dra I) and the ultra-faint Coma Berenices (CBe), Canes Venatici I (CVn I), Segue 1 (Seg 1), Reticulum II (Ret II) and Triangulum II (Tri II). With respect to previous works, all of the results shown here are obtained with a common methodology for all targets, and will be included, along with other promising dSphs, in a future comprehensive publication on this research topic.

\begin{figure*}[ht]
\centering
\includegraphics[scale=0.11,angle=-90]{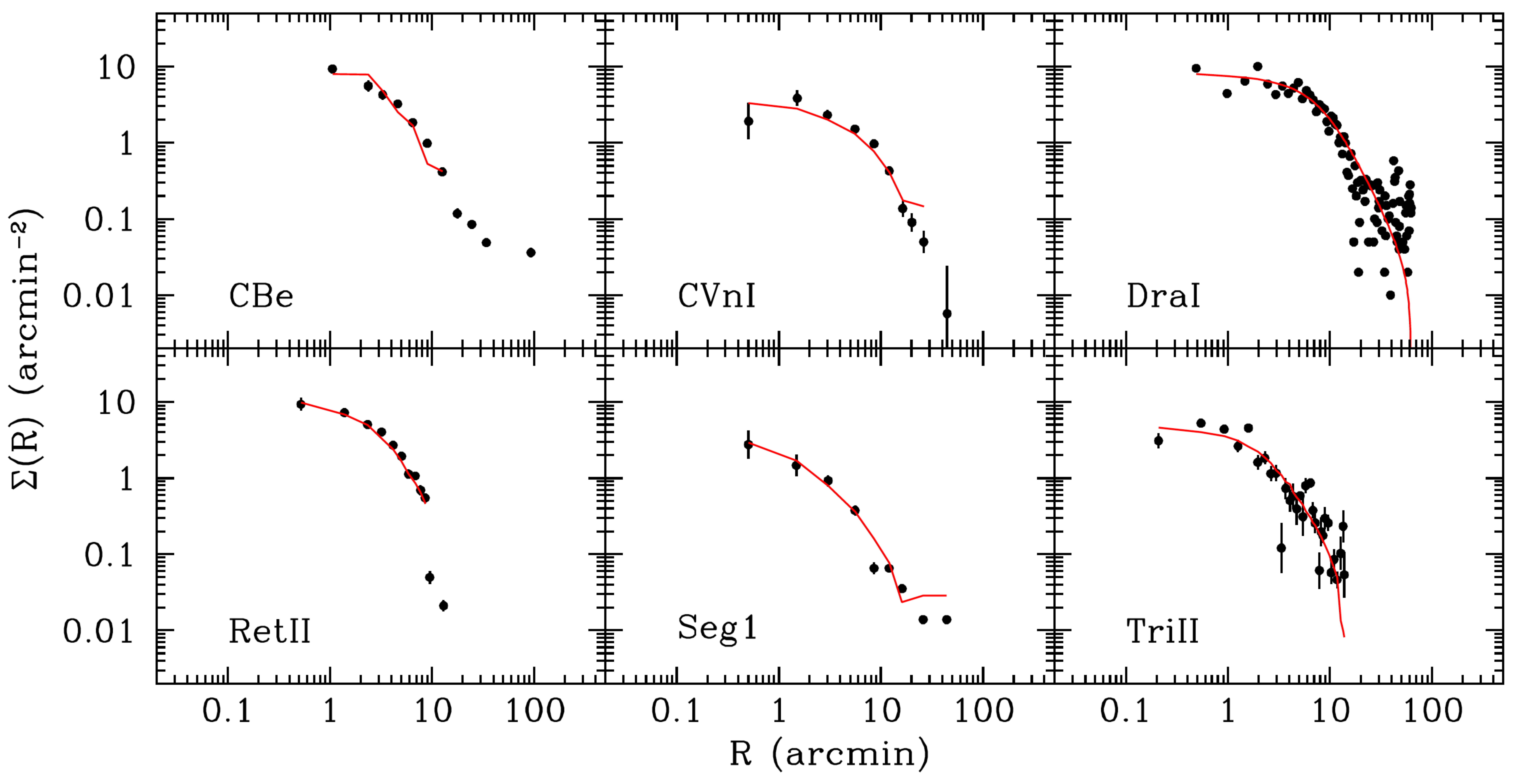}
\caption{\footnotesize{Best-fit brightness profiles $\Sigma(R)$ of the analyzed dSphs as a function of the object's projected (2D) radial coordinate $R$ from the dSph centroid. In each panel, the projected 3D profile resulting from the fit ({\itshape red line}) is shown superimposed to the corresponding data set ({\itshape black dots}; \cite{Irw95,Mar08,Mun10,Bec15,Lae15}).}}
\label{fig-1}
\end{figure*}

\section{Indirect DM detection in the $\gamma$-ray domain}
\label{sec-1}

The $\gamma$-ray flux expected from DM self-interaction depends on the type of DM reaction, the mass of the DM particle and the amount of DM along the line of sight. In case of annihilation of two DM particles, the latter quantity is summarized into the astrophysical factors for DM annihilation $J(\Delta \Omega)$ or decay $D(\Delta\Omega)$:
\begin{eqnarray}\label{eqn:jfac}
J(\Delta \Omega) = \int_{\Delta \Omega} d\Omega \int_{\rm l.o.s.} \rho_{\rm DM}^2(\ell, \Omega) d\ell\\
D(\Delta \Omega) = \int_{\Delta \Omega} d\Omega \int_{\rm l.o.s.} \rho_{\rm DM}(\ell, \Omega) d\ell
\end{eqnarray}
Several attempts to get such values with increasing accuracy exist in the literature \citep{Acc10,Bon15,Ger15,Mar15,Hay16}. Here we present new determinations of the astrophysical factors for six of the best dSphs known to date in the field of DM indirect searches, based on the procedure described in \cite{Bon15} which exploits the capabilities of the Markov-Chain Monte Carlo (MCMC) Jeans analysis performed by the {\scriptsize CLUMPY} code \cite{Hut18}. At variance with them, we adopt a common framework for the treatment of surface-brightness and kinematic data sets of the analyzed dSphs, including the most promising newly-discovered targets.

\begin{table*}
\caption{\footnotesize{{\itshape Top:} best-fit parameters -- heliocentric distance $d_\odot$ \cite{McC12}; 3D brightness scale density $\rho^*_{\rm s}$, radius $r^*_{\rm s}$ and profile indexes; dSph virial radius $R_{\rm vir}$; 3D DM density scale $\rho_{\rm s}$, radius $r_{\rm s}$ and index $\alpha$ \cite{Bon15} -- for the six analyzed dSphs, classified for type (`cls': classical, `uft': ultra-faint). {\itshape Bottom:} logarithmic astrophysical factors $\log{J}$ and $\log{D}$ for DM self-interaction, computed at their optimal angles ($\alpha_J = 2r_{\rm s}^*/d_\odot$ or $\alpha_D = r_{\rm s}^*/d_\odot$ respectively) and at $0.5$ deg for a direct comparison with the literature \cite{Bon15,Bon15c,Gen16}.}}
\label{tab-1}
\begin{center}
\resizebox{.82\textwidth}{!}{
\begin{tabular}{lccccccccccc}
\hline
\hline
\addlinespace[1ex] Name & Type & $d_\odot$ & $\rho_{\rm s}^*$ & $r_{\rm s}^*$ & $\alpha^*$ & $\beta^*$ & $\gamma^*$ & $R_{\rm vir}$ & $\rho_{\rm s}$ & $r_{\rm s}$ & $\alpha$\\
 & & (kpc) & ($10^5$ L$_\odot$ kpc$^{-3}$) & (kpc) & & & & (kpc) & ($10^6$ M$_\odot$ kpc$^{-3}$) & (kpc) & \\
 \addlinespace[1ex] \hline
\addlinespace[1ex] CBe & uft & $44 \pm 4$ & $3.7 \pm 1.7$ & $0.092 \pm 0.008$ & 3.8 & 3.8 & 1.4 & $4.2^{+14.5}_{-2.7}$ & $37^{+54}_{-28}$ & $1.0^{+4.6}_{-0.7}$ & $0.64^{+0.25}_{-0.29}$ \\
CVn I & uft & $218 \pm 10$ & $2.15 \pm 0.50$ & $0.568 \pm 0.033$ & 3.2 & 3.8 & 0.2 & $12.2^{+3.2}_{-4.0}$ & $2.5^{+6.7}_{-1.4}$ & $1.9^{+2.8}_{-1.1}$ & $0.32^{+0.32}_{-0.45}$ \\
Dra I & cls & $76 \pm 6$ & $470 \pm 140$ & $0.170 \pm 0.016$ & 2.8 & 3.8 & 0.2 & $2.68^{+0.45}_{-0.32}$ & $85^{+68}_{-29}$ & $0.28^{+0.33}_{-0.20}$ & $0.41^{+0.23}_{-0.21}$ \\
Ret II & uft & $30 \pm 3$ & $157 \pm 24$ & $0.023 \pm 0.003$ & 3.2 & 3.8 & 0.4 & $1.7^{+5.9}_{-1.1}$ & $65^{+220}_{-50}$ & $0.41^{+2.36}_{-0.34}$ & $0.56^{+0.29}_{-0.27}$ \\
Seg 1 & uft & $23 \pm 2$ & $13 \pm 11$ & $0.034 \pm 0.007$ & 3.6 & 3.8 & 0.8 & $0.28^{+1.94}_{-0.23}$ & $1.6^{+22.7}_{-1.3}$ & $0.32^{+4.02}_{-0.25}$ & $0.53^{+0.33}_{-0.26}$ \\
Tri II & uft & $30 \pm 2$ & $45 \pm 24$ & $0.020 \pm 0.002$ & 3.6 & 3.2 & 0.2 & $6.2^{+15.0}_{-4.7}$ & $800^{+2070}_{-660}$ & $0.64^{+1.76}_{-0.55}$ & $0.52^{+0.32}_{-0.27}$ \\
\addlinespace[1ex] \hline
\addlinespace[1ex] & $\alpha_J$ & \multicolumn{2}{c}{$\log{J(\alpha_J)}$} & \multicolumn{2}{c}{$\log{J(0.5\mbox{ deg})}$} & \multicolumn{2}{c}{$\alpha_D$} & \multicolumn{2}{c}{$\log{D(\alpha_D)}$} & \multicolumn{2}{c}{$\log{D(0.5\mbox{ deg})}$} \\
 & (deg) & \multicolumn{4}{c}{(GeV$^2$ cm$^{-5}$)} & \multicolumn{2}{c}{(deg)} & \multicolumn{4}{c}{(GeV cm$^{-2}$)} \\
 \addlinespace[1ex] \hline
\addlinespace[1ex] CBe & 0.24 & \multicolumn{2}{c}{$19.1^{+0.7}_{-0.6}$} & \multicolumn{2}{c}{$19.4^{+0.9}_{-0.7}$} & \multicolumn{2}{c}{0.12} & \multicolumn{2}{c}{$17.7^{+0.6}_{-0.4}$} & \multicolumn{2}{c}{$18.7^{+0.7}_{-0.6}$} \\
CVn I & 0.30 & \multicolumn{2}{c}{$17.1^{+0.2}_{-0.2}$} & \multicolumn{2}{c}{$17.2^{+0.2}_{-0.2}$} & \multicolumn{2}{c}{0.15} & \multicolumn{2}{c}{$17.0^{+0.1}_{-0.1}$} & \multicolumn{2}{c}{$17.6^{+0.2}_{-0.2}$} \\
Dra I & 0.26 & \multicolumn{2}{c}{$18.6^{+0.1}_{-0.2}$} & \multicolumn{2}{c}{$18.6^{+0.1}_{-0.1}$} & \multicolumn{2}{c}{0.13} & \multicolumn{2}{c}{$17.4^{+0.1}_{-0.1}$} & \multicolumn{2}{c}{$18.0^{+0.1}_{-0.1}$} \\
Ret II & 0.09 & \multicolumn{2}{c}{$18.6^{+0.5}_{-0.5}$} & \multicolumn{2}{c}{$19.3^{+0.9}_{-0.7}$} & \multicolumn{2}{c}{0.05} & \multicolumn{2}{c}{$16.9^{+0.4}_{-0.3}$} & \multicolumn{2}{c}{$18.5^{+0.7}_{-1.0}$} \\
Seg 1 & 0.17 & \multicolumn{2}{c}{$16.3^{+2.1}_{-2.1}$} & \multicolumn{2}{c}{$16.7^{+2.3}_{-2.5}$} & \multicolumn{2}{c}{0.09} & \multicolumn{2}{c}{$16.0^{+1.2}_{-1.2}$} & \multicolumn{2}{c}{$17.0^{+1.4}_{-2.2}$} \\
Tri II & 0.08 & \multicolumn{2}{c}{$20.8^{+0.6}_{-0.5}$} & \multicolumn{2}{c}{$21.6^{+1.0}_{-0.9}$} & \multicolumn{2}{c}{0.04} & \multicolumn{2}{c}{$17.8^{+0.5}_{-0.4}$} & \multicolumn{2}{c}{$19.6^{+0.8}_{-1.0}$} \\
\addlinespace[1ex] \hline
\end{tabular}
}
\end{center}
\end{table*}

\subsection{Input priors and data sets for the {\scriptsize CLUMPY} set-up}

The {\scriptsize CLUMPY} software allows to perform the dynamical analysis of the DM halos around dSphs assuming that such galaxies can be considered steady-state collisionless systems in spherical symmetry and with negligible rotation, in which the contribution of the stellar component to the total mass can be neglected. In these conditions, their second-order Jeans equation \cite{Bin08} can be solved to obtain the DM density profile $\rho_{\rm DM}(r)$ along the object's 3D radial coordinate $r$, once parametric forms of the stellar number density, radial velocity dispersion and velocity anisotropy are given. We follow the prescription by \cite{Bon15b}, fitting the brightness density $n^*(r)$ of each galaxy with a 3D Navarro-Frenk-White (NFW) profile \cite{Nav97} projected onto the corresponding surface brightness (see Fig. \ref{fig-1} and Tab. \ref{tab-1}) and adopting an Einasto profile \cite{Ein65} for $\rho_{\rm DM}$, so as to be directly comparable with the majority of the values of $J$ and $D$ available in the literature. For the dSph stellar kinematics, we use the same data sets of \cite{Bon15}, \cite{Bon15c} and \cite{Gen16}, also adopting the definition of stellar membership probabilities $P$ by \cite{Bon15}.

\subsection{Astrophysical factors from the MCMC spherical Jeans analysis of the dSph halos}

\begin{figure*}[ht]
\centering
\includegraphics[scale=0.11,angle=-90]{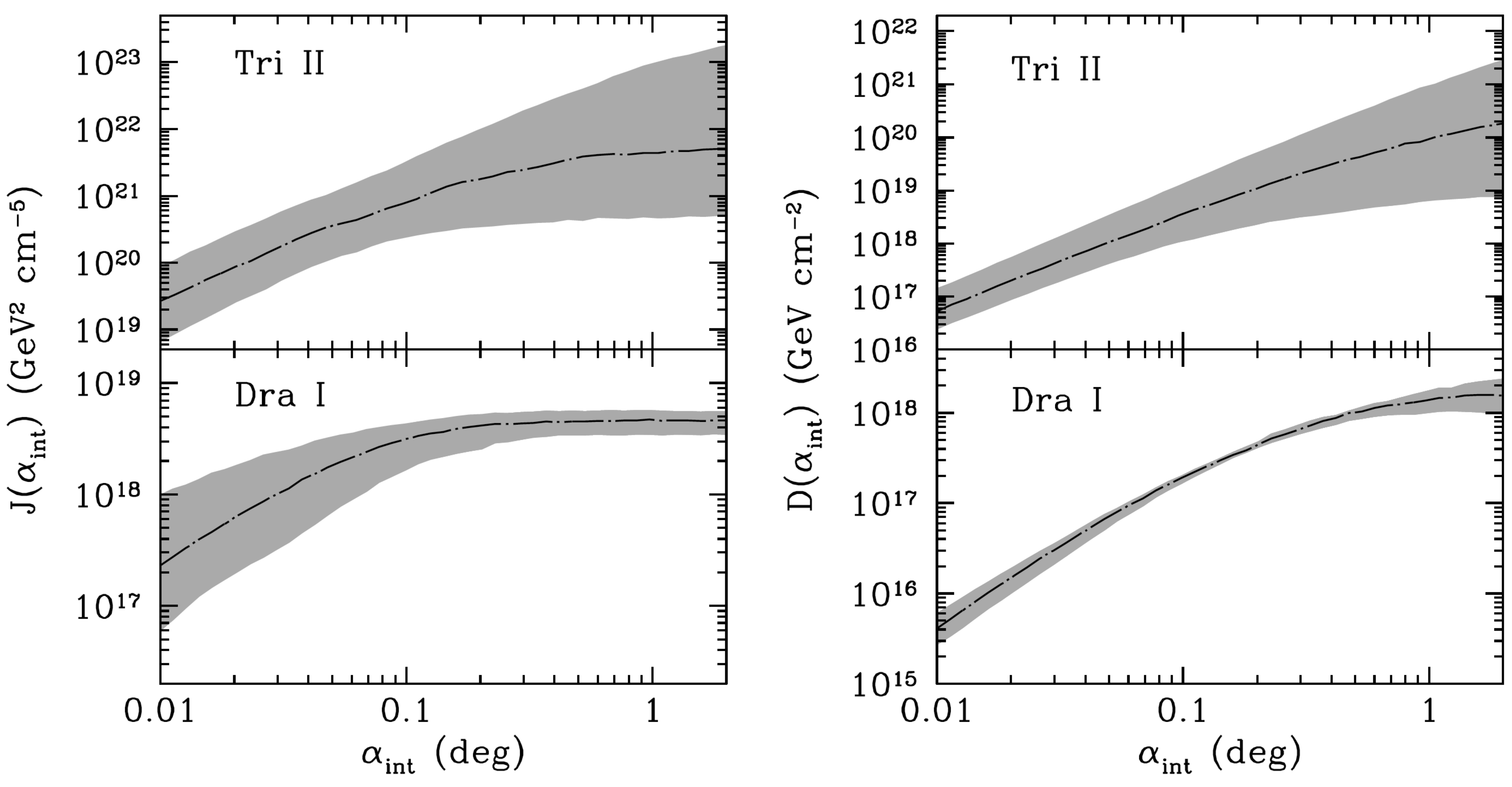}
\caption{\footnotesize{{\itshape Left panels:} DM-annihilation astrophysical factor profiles $J(\alpha_{\rm int})$ as functions of the integration angle $\alpha_{\rm int}$ for the most massive halo (Tri II) and the halo with the most robust parameter determination (Dra I). {\itshape Right panels:} equivalent profiles $D(\alpha_{\rm int})$ for DM decay.}}
\label{fig-2}
\end{figure*}

We run 10 MC chains with $10^4$ realizations each for every target, choosing the unbinned likelihood analysis \cite{Bon15b}. We then derive the distribution of virial radii $R_{\rm vir}$ for each galaxy from the output profiles as made by \cite{Bon15}. Finally, we compute the profiles of $J(\alpha_{\rm int})$ and $D(\alpha_{\rm int})$ as functions of the integration angle $\alpha_{\rm int}$ by running the {\scriptsize CLUMPY} executable over the posterior distributions of the DM profile parameters. In Tab. \ref{tab-1}, we report the values of the astrophysical factors obtained in this way for different integration angles, in order to favor a direct comparison with the results already available in the literature. In Fig. \ref{fig-2}, we also show the profiles of $J$ and $D$ for the dSph with the highest expected DM amount (Tri II) and that with the most robust DM halo parameters (Dra I).

\section{Discussion}

Our MCMC examination of the first six DM halos in a completely uniform framework of input data allows us to produce $J$ and $D$ values for dSphs in agreement with the literature. The most discrepant result is $J(0.5\mbox{ deg})$ for Dra I, which is offset by a factor of $\sim$4 from e.g. the value given by \cite{Bon15}. Nevertheless, {\scriptsize CLUMPY} is only able to perform a {\itshape spherical} Jeans analysis of the DM halos, thus not taking into account additional uncertainties arising from a possible halo triaxiality ($\sim$0.4 dex). We also note that, within the analyzed data sets, the dSph Tri II is potentially observable with CTA as a single targets in $\sim$100 h integration. However, the limited statistics available on its stellar kinematics (which is also possibly contaminated by binary stars \cite{Kir17}) prevents to draw any firm conclusion, again emphasizing the importance of collecting good and clean stellar data for such targets in order to evaluate them as viable candidates for DM searches with Cherenkov telescopes.

\bibliography{biblio.bib}

\end{document}